\documentstyle[prl,aps,multicol]{revtex}

\begin{document}
\draft
\title{
{\Large \bf Phase Transitions of the Flux Line Lattice in
High-Temperature Superconductors with Weak Columnar and Point
Disorder} }
\author{Yadin Y. Goldschmidt}
\address{Department of Physics and Astronomy\\
University of Pittsburgh\\
Pittsburgh, PA 15260}
\date{August 4, 1996}
\maketitle

\begin{abstract}
We study the effects of weak columnar and point disorder
on the vortex-lattice phase transitions in high temperature
superconductors. The combined effect of thermal
fluctuations and of quenched disorder is investigated using
a simplified cage model. For columnar disorder the problem maps into a
quantum particle in a harmonic + random potential. We use the
variational approximation to show that columnar and point disorder
have opposite effect on the position of the melting line as observed
experimentally. Replica symmetry breaking plays a role at the
transition into a vortex glass at low temperatures.

\end{abstract}


\begin{multicols}{2}

There is a lot of interest in the physics of high temperature
superconductors due to their potential technological applications. In
particular these materials are of type II and allow for partial magnetic
flux penetration. Pinning of the magnetic flux lines (FL) by many types of
disorder is essential to eliminate dissipative losses associated with flux
motion. In clean materials below the superconducting temperature there exist
a 'solid ' phase where the vortex lines form a triangular Abrikosov lattice 
\cite{blatter}. This solid can melt due to thermal fluctuations and the
effect of impurities. In particular known observed transitions are into a
flux liquid at higher temperatures via a {\it melting line} (ML)\cite{zeldov},
and into a vortex glass at low temperature \cite{VG},\cite{Fisher},\cite{BG}
in the presence of disorder- the so called {\it entanglement line} (EL). \cite
{blatter}

Recently the effect of point and columnar disorder on the position of the
melting transition has been measured experimentally in the high-$T_c$
material $Bi_2Sr_2CaCu_2O_8$ \cite{Khaykovitch}. Point disorder has been
induced by electron irradiation (with 2.5 MeV electrons), whereas columnar
disorder has been induced by heavy ion irradiation (1 GeV Xe or 0.9 GeV Pb).
It turns out that the flux melting transition persists in the presence of
either type of disorder, but its position shifts depending on the disorder
type and strength.

A significant difference has been observed between the effects of columnar
and point disorder on the location of the ML. Weak columnar defects stabilize
the solid phase with respect to the vortex liquid phase and shift the
transition to {\it higher} fields, whereas point-like disorder destabilizes
the vortex lattice and shifts the melting transition to {\it lower} fields.
In this paper we attempt to provide an explanation to this observation. The
case of point defects has been addressed in a recent paper by Ertas and
Nelson \cite{EN} using the cage-model approach which replaces the effect of
vortex-vortex interactions by an harmonic potential felt by a single vortex.
For columnar disorder the parabolic cage model was introduced by
Nelson and Vinokur [8]. Here we use a different approach to analyze
the cage-model Hamiltonian vis. the replica method together with the
variational approximation. In the case of columnar defects our
approach relies on our recent analysis of a quantum particle in a
random potential \cite{yygold}. We compare the effect of the two types
of disorder with each other and with results of recent experiments.

Assume that the average magnetic field is aligned along the $z$-axis.
Following EN we describe the Hamiltonian of a single FL whose position is
given by a two-component vector ${\bf r}(z)$ (overhangs are neglected) by: 
\begin{eqnarray}
{\cal H} = \int_0^L dz \left\{ {\frac{\tilde{\epsilon} }{2}} \left({\frac{ d%
{\bf r }}{dz}} \right)^2 + V(z,{\bf r }) + {\frac{\mu }{2}} {\bf r }^2
\right\}.  \label{hamil}
\end{eqnarray}

Here $\tilde \epsilon =\epsilon _0/\gamma ^2$ is the line tension of the FL, 
$\gamma ^2=m_z/m_{\perp }$ is the mass anisotropy, $\epsilon _0=(\Phi
_0/4\pi \lambda )^2$, ($\lambda $ being the penetration length), and $\mu
\approx \epsilon _0/a_0^2$ is the effective spring constant (setting the
cage size) due to interactions with neighboring FLs, which are at a typical
distance of $a_0=\sqrt{\Phi _0/B}$ apart.

For the case of columnar (or correlated) disorder, $V(z,{\bf r})=V({\bf r})$
is independent of $z$, and 
\begin{eqnarray}
\langle V({\bf r})V({\bf r^{\prime }})\rangle \equiv -2f(({\bf r}-{\bf %
r^{\prime }})^2/2)=g\epsilon _0^2\xi ^2\delta _\xi ^{(2)}({\bf r}-{\bf %
r^{\prime }}),  \label{VVC}
\end{eqnarray}
where 
\begin{eqnarray}
\delta _\xi ^{(2)}({\bf r}-{\bf r^{\prime }})\approx 1/(2\pi \xi ^2)\exp (-(%
{\bf r}-{\bf r^{\prime }})^2/2\xi ^2),  \label{delta}
\end{eqnarray}
and $\xi $ is the vortex core diameter. The dimensionless parameter g is a
measure of the strength of the disorder. On the other hand for
point-disorder, $V$ depends on $z$ and \cite{EN} 
\begin{eqnarray}
\langle V(z,{\bf r})V(z^{\prime },{\bf r^{\prime }})\rangle =\tilde
\Delta \epsilon 
_0^2\xi ^3\delta _\xi ^{(2)}({\bf r}-{\bf r^{\prime }})\delta (z-z^{\prime
}).  \label{VVP}
\end{eqnarray}

The quantity that measures the transverse excursion of the FL is 
\begin{eqnarray}
u_0^2(\ell )\equiv \langle |{\bf r}(z)-{\bf r}(z+\ell )|^2\rangle \ /2,
\label{ul}
\end{eqnarray}

Let us now review the connection between a quantum particle in a random
potential and the behavior of a FL in a superconductor. The partition
function of the former is just like the partition sum of the FL, provided
one make the identification \cite{nelson} 
\begin{eqnarray}
\hbar \rightarrow T,\qquad \beta \hbar \rightarrow L,  \label{corresp}
\end{eqnarray}
Where T is the temperature of the superconductor and L is the system size in
the $z$-direction. $\beta $ is the inverse temperature of the quantum
particle. We are interested in large fixed L as T is varied, which
corresponds to high $\beta $ for the quantum particle when $\hbar $ (or
alternatively the mass of the particle) is varied. The variable $z$ is the
so called Trotter time. This is the picture we will be using for the case of
columnar disorder.

For the case of point-disorder the picture we use is that of a directed
polymer in the presence of a random potential plus an harmonic potential as
used by EN.

The main effect of the harmonic (or cage) potential is to cap the transverse
excursions of the FL beyond a confinement length $\ell ^{*}\approx
a_0/\gamma $. The mean square displacement of the flux line is given by

\begin{equation}
u^2(T)\approx u_0^2(\ell ^{*}).  \label{uT}
\end{equation}

The location of the melting line is determined by the Lindemann criterion
\begin{equation}
u^2(T_m(B))=c_L^2a_0^2,  \label{Lind}
\end{equation}
where $c_L\approx 0.15-0.2$ is the phenomenological Lindemann constant. This
means that when the transverse excursion of a section of length $\approx
\ell ^{*}$becomes comparable to a finite fraction of the interline
separation $a_0$, the melting of the flux solid occurs.

We consider first the case of columnar disorder. In the absence of disorder
it is easily obtained from standard quantum mechanics and the correspondence
(\ref{corresp}), that when $L\rightarrow \infty ,$

\begin{equation}
u^2(T)=\frac T{\sqrt{\widetilde{\epsilon }\mu }}\left( 1-\exp (-\ell ^{*}%
\sqrt{\mu /\widetilde{\epsilon }})\right) =\frac T{\sqrt{\widetilde{\epsilon 
}\mu }}(1-e^{-1}),  \label{u2g0}
\end{equation}
from which we find that

\begin{equation}
B_m(T)\approx \frac{\Phi _0^{}}{\xi ^2}\frac{\epsilon _0^2\xi ^2c_L^4}{%
\gamma ^2T^2}.  \label{Bmg0}
\end{equation}

When we turn on disorder we have to solve the problem of a quantum particle
in a random quenched potential. This problem has been recently solved using
the replica method and the variational approximation \cite{yygold}. Let us
review briefly the results of this approach. In this approximation we chose
the best quadratic Hamiltonian parametrized by the matrix $%
s_{ab}(z-z^{\prime })$:

\begin{eqnarray}
h_n &=&\frac 12\int_0^Ldz\sum_a[\widetilde{\epsilon }{\bf \dot r}_a^2+\mu 
{\bf r}_a^2]  \nonumber \\
&&-\frac 1{2T}\int_0^Ldz\int_0^Ldz^{\prime }\sum_{a,b}s_{ab}(z-z^{\prime })%
{\bf r}_a(z)\cdot {\bf r}_b(z^{\prime }).  \label{hn}
\end{eqnarray}
Here the replica index $a=1\ldots n$, and $n\rightarrow 0$ at the end of the
calculation. This Hamiltonian is determined by stationarity of the
variational free energy which is given by

\begin{equation}
\left\langle F\right\rangle _R/T=\left\langle H_n-h_n\right\rangle
_{h_n}-\ln \int [d{\bf r}]\exp (-h_n/T),  \label{FV}
\end{equation}
where $H_n$ is the exact $n$-body replicated Hamiltonian. The off-diagonal
elements of $s_{ab}$can consistently be taken to be independent of $z$,
whereas the diagonal elements are $z$-dependent. It is more convenient to
work in frequency space, where $\omega $ is the frequency conjugate to $z$. $%
\omega _j=(2\pi /L)j,$with $j=0,\pm 1,\pm 2,\ldots $.Assuming replica
symmetry, which is valid only for part of the temperature range, we can
denote the off-diagonal elements of $\widetilde{s}_{ab}(\omega
)=(1/T)\int_0^Ldz\ e^{i\omega z}$ $s_{ab}(z)$, by $\widetilde{s}(\omega )=%
\widetilde{s}\delta _{\omega ,0}$. Denoting the diagonal elements by $%
\widetilde{s}_d(\omega )$, the variational equations become: 
\begin{eqnarray}
\tilde s &=&2\frac LT\widehat{f}\ ^{\prime }\left( {\frac{2T}{\mu L}}+{\frac{%
2T}L}\sum_{\omega ^{\prime }\neq 0}\frac 1{\epsilon \ \omega ^{\prime
}\,^2+\mu -\widetilde{s}_d(\omega ^{\prime })}\right)  \label{s} \\
\tilde s_d(\omega ) &=&\tilde s-{\frac 2T}\int_0^Ld\zeta \ (1-e^{i\omega
\zeta })\times  \nonumber \\
&&\ \ \widehat{f}\ ^{\prime }\left( {\frac{2T}L}\sum_{\omega ^{\prime }\neq
0}\ \frac{1-e^{-i\omega ^{\prime }\varsigma }}{\widetilde{\epsilon \ }\omega
^{\prime }\,^2+\mu -\widetilde{s}_d(\omega ^{\prime })}^{}\right) .
\label{sd}
\end{eqnarray}
here $\widehat{f}$ $^{\prime }(y)$ denotes the derivative of the ''dressed''
function $\widehat{f}(y)$ which is obtained in the variational scheme from
the random potential's correlation function $f(y)$ (see eq. (\ref{VVC})),
and in 2+1 dimensions is given by:

\begin{equation}
\widehat{f}(y)=-\frac{g\epsilon _0^2\xi ^2}{4\pi }\frac 1{\xi ^2+y}
\label{f}
\end{equation}
The full equations, taking into account the possibility of replica-symmetry
breaking are given in ref. \cite{yygold}. In terms of the variational
parameters the function $u_0^2(\ell ^{*})$ is given by

\begin{equation}
u_0^2(\ell ^{*})={\frac{2T}L}\sum_{\omega ^{\prime }\neq 0}\frac{1-\cos
(\omega ^{\prime }\ell ^{*})}{\widetilde{\epsilon \ }\omega ^{\prime
}\,^2+\mu -\widetilde{s}_d(\omega ^{\prime })}.  \label{u2qp}
\end{equation}
This quantity has not been calculated in ref. \cite{yygold}. There we
calculated $\left\langle {\bf r}^2(0)\right\rangle $ which does not measure
correlations along the $z$-direction.

In the limit $L\rightarrow \infty $ we were able to solve the equations
analytically to leading order in $g$. In that limit eq. (\ref{sd}) becomes
(for $\omega \neq 0$) :

\begin{eqnarray}
\tilde s_d(\omega ) &=&\frac 4\mu \widehat{f}\ ^{\prime \prime }(b_0)-\frac 2%
T\int_0^\infty d\varsigma (1-\cos (\omega \varsigma ))  \nonumber \\
&&\times (\widehat{f}\ ^{\prime }(C_0(\varsigma ))-\widehat{f}\ ^{\prime
}(b_0)),  \label{sdi}
\end{eqnarray}
with

\begin{equation}
C_0(\varsigma )=2T\int_{-\infty }^\infty \frac{d\omega }{2\pi }\frac{1-\cos
(\omega \varsigma )}{\widetilde{\epsilon \ }\omega \,^2+\mu -\widetilde{s}%
_d(\omega )}  \label{C0}
\end{equation}
and $b_0$ given by a similar expression with the cosine term missing in the
numerator of eq. (\ref{C0}).

Defining

\begin{eqnarray}
\tau &=&T\ /\sqrt{\widetilde{\epsilon }\ \mu },\ \alpha =\tau \ /(\xi
^2+\tau ),  \label{tau,al} \\
f_1(\alpha ) &=&1/(1-\alpha )-(1/\alpha )\log (1-\alpha ),  \label{f1} \\
f_2(\alpha ) &=&\frac 1\alpha \sum_{k=1}^\infty (k+1)\alpha ^k/k^3
\label{f2} \\
a^2 &=&f_1(\alpha )/f_2(\alpha ),\ A=-\widehat{f}\ ^{\prime \prime }(\tau )\
f_1^2(\alpha )/f_2(\alpha )/\mu ,  \label{a2,A} \\
s_\infty &=&\widehat{f}\ ^{\prime \prime }(\tau )\ (4+f_1(\alpha ))/\mu ,
\label{sinf}
\end{eqnarray}
a good representation of $\widetilde{s}_d(\omega ),\ (\omega \neq 0)$ with
the correct behavior at low and high frequencies is

\begin{equation}
\widetilde{s}_d(\omega )=s_\infty +A\mu /(\widetilde{\epsilon \ }\omega
^2+a^2\mu ).  \label{sde}
\end{equation}
(notice that this function is negative for all $\omega $). Substituting in
eq. (\ref{C0}) and expanding the denominator to leading order in the
strength of the disorder, we get :

\begin{eqnarray}
u_0^2(\ell ) &=&C_0(\sqrt{\widetilde{\epsilon }\ /\ \mu })=\tau
(1-A/(a^2-1)^2/\mu )  \nonumber \\
&&\ \times (1-e^{-\ell /\ell ^{*}})+\tau A/(a(a^2-1)^2\mu )\times  \nonumber
\\
&&(1-e^{-a\ell /\ell ^{*}})+\tau /(2\mu )\times \ (s_\infty +A/(a^2-1)) 
\nonumber \\
&&\times \ (1-e^{-\ell /\ell ^{*}}-(\ell /\ell ^{*})\ e^{-\ell /\ell ^{*}}).
\label{u2f}
\end{eqnarray}
In order to plot the results we measure all distances in units of $\xi $ ,
we measure the temperature in units of $\epsilon _0\xi $, and the magnetic
field in units of $\Phi _0/\xi ^2$ . We observe that the spring constant $%
\mu $ is given in the rescaled units by $B$ and $a_0=1/\sqrt{B}$. We further
use $\gamma =1$ for the plots.

Fig. 1 shows a plot of $\sqrt{u_0^2(\ell ^{*})}/a_0$ vs. $T$ for zero disorder
(curve a) as well as for $g/2\pi =0.02$ (curve b). We have chosen $B=1/900$.
We see that the disorder tends to align the flux lines along the columnar
defects , hence decreasing $u^2(T)$ .Technically this happens since $%
\widetilde{s}_d(\omega )$ is negative. The horizontal line represents a
possible Lindemann constant of 0.15.

In Fig. 2 we show the modified melting line $B_m(T)$ in the presence of
columnar disorder. This is obtained from eq. (\ref{Lind}) with $c_L=0.15$.
We see that it shifts towards higher magnetic fields.

For $T<T_c\approx (\epsilon _0\xi /\gamma )[g^2\epsilon _0/(16\pi ^2\mu \xi
^2)]^{1/6}$, there is a solution with RSB but we will not pursue it further
in this paper. This temperature is at the bottom of the range plotted in the
figures for columnar disorder. We will pursue the RSB solution only for the
case of point disorder, see below. The expression (\ref{u2f}) becomes
negative for very low temperature. This is an artifact of the
truncation of the expansion in the strength of the disorder.

For the case of point defects the problem is equivalent to a directed
polymer in a combination of a random potential and a fixed harmonic
potential. This problem has been investigated by MP \cite{mp}, who were
mainly concerned with the limit of $\mu \rightarrow 0$. In this case the
variational quadratic Hamiltonian is parametrized by:

\begin{eqnarray}
h_n &=&\frac 12\int_0^Ldz\sum_a[\widetilde{\epsilon }{\bf \dot r}_a^2+\mu 
{\bf r}_a^2]  \nonumber \\
&&\ \ -\frac 12\int_0^Ldz\sum_{a,b}^{}s_{ab}\ {\bf r}_a(z)\cdot {\bf r}_b(z),
\label{hnpd}
\end{eqnarray}
with the elements of $s_{ab}$ all constants as opposed to the case of
columnar disorder.

The replica symmetric solution to the variational equations is simply given
by :

\begin{eqnarray}
s &=&s_d=\frac{2\xi }T\widehat{f}\ ^{\prime }(\tau )  \label{s,sd} \\
u_0^2(\ell ) &=&2T\int_{-\infty }^\infty \frac{d\omega }{2\pi }\frac{1-\cos
(\omega \ell )}{\widetilde{\epsilon \ }\omega \,^2+\mu } \left( 1+
\frac{s_d}{ \widetilde{\epsilon \ }\omega \,^2+\mu}\right) \label{u2p}
\end{eqnarray}
and hence

\begin{eqnarray}
u_0^2(\ell ) &=&\tau (1-e^{-\ell /\ell ^{*}})+\tau \ s_d\ /\ (2\mu )  \nonumber
\\
&&\ \ \times \ (1-e^{-\ell /\ell ^{*}}-(\ell /\ell ^{*})\ e^{-\ell /\ell
^{*}}).  \label{u2p2}
\end{eqnarray}

In eq.(\ref{s,sd}) $\widehat{f}$ is the same function as defined in eq. (\ref
{f}) with $g\ $replaced by $\widetilde{\Delta }$. As opposed the case of
columnar disorder, in this case $s_d$ is positive and independent of $\omega 
$, and hence the mean square displacement $u_0^2(\ell ^{*})$ is bigger than
its value for zero disorder. Fig. 1 curve {\it c }shows a plot of $\sqrt{%
u_0^2(\ell ^{*})}/a_0$ vs. $T$ for $\widetilde{\Delta }/2\pi =0.8$. Again $%
B=1/900$. For $T<T_{cp}\approx $ $(\epsilon _0\xi /\gamma )(\gamma $ $%
\widetilde{\Delta }/2\pi )^{1/3}$ it is necessary to break replica symmetry
as shown by MP \cite{mp}. This means that the off-diagonal elements of the
variational matrix $s_{ab}$ are not all equal to each other. MP worked out
the solution in the limit of $\mu \rightarrow 0$, but it is not difficult to
extend it to any value of $\mu .$ We have worked out the first stage RSB
solution which is all is required for a random potential with short ranged
correlations. The analytical expression is not shown here for lack of space.
The solution is represented by curve {\it d} in Fig. 1 which consists of
upward triangles.

The modified melting line in the presence of disorder is indicated by the
curve {\it c} in Fig. 2 for $T>T_{cp}$. For $T<T_{cp}$ the so called {\it %
entanglement line }is represented by curve {\it d} of filled squares.The
value of the magnetic field $B_m(T_{cp})\approx (\Phi _0/\xi ^2)(\gamma 
\widetilde{\Delta }/2\pi )^{-2/3}c_L^4$  gives a reasonable agreement
with the experiments.

The analytical expressions given in eqs. (\ref{u2f}), (\ref{u2p2}), though
quite simple, seem to capture the essential feature required to reproduce
the position of the melting line. The qualitative agreement with
experimental results is remarkable, especially the opposite effects of
columnar and point disorder on the position of the melting line. The 'as
grown' experimental results are corresponding to very small amount of point
disorder, and thus close to the line of no disorder in the figures. At low
temperature, the entanglement transition is associated in our formalism with
RSB, and is a sort of a spin-glass transition in the sense that many minima
of the random potential and hence free energy, compete with each other. In
this paper we worked out the one-step RSB for the case of point disorder.
The experiments show that in the case of colmunar
disorder the transition into the vortex glass seems to be absent. This has
to be further clarified theoretically. We have shown that the {\it cage
model }together with the variational approximation reproduce the main
feature of the experiments. Effects of many body interaction between vortex
lines which are not taken into account by the effective cage model seem to
be of secondary importance. Inclusion of such effects within the variational
formalism remains a task for the future.

For point disorder, in the limit of infinite cage ( $\mu \rightarrow
0$), the variational approximation gives a wandering exponent of 1/2
for a random potential with short ranged correlations \cite{mp},
whereas simulations give a value of 5/8 \cite{halpin}. This
discrepancy does not seem of importance with respect to the conclusions
obtained in this paper. Another point 
to notice is that columnar disorder is much more effective in shifting the
position of the melting line as compared for point disorder in the range of
parameters considered here. We have used a much weaker value of correlated
disorder to achieve a similar or even larger shift of the melting line than
for the case of point disorder. The fact that the
random potential does not vary along the z-axis enhances its effect on the
vortex lines.

We thank David Nelson and Eli Zeldov for discussions. We thank the Weizmann
institute for a Michael Visiting Professorship, during which this research
has been carried out.

Figure Captions: Fig1: Transverse fluctuations in the cage model for (a) no
disorder (b)columnar disorder (c)point disorder (d)RSB for point disorder.
Fig. 2: Melting line for (a) no disorder (b) columnar disorder (c)point
disorder (d) entanglement line for point disorder.

\end{multicols}
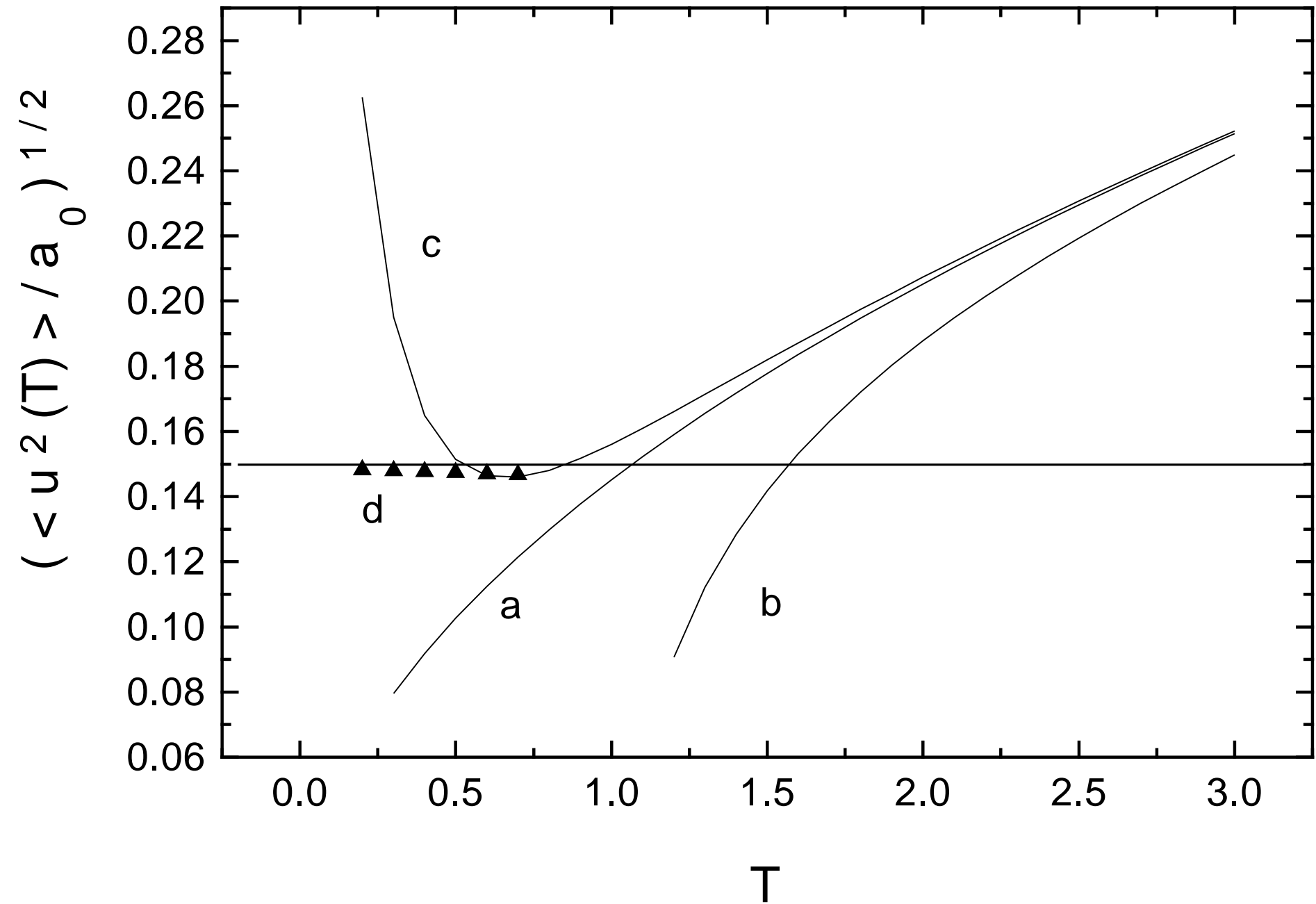
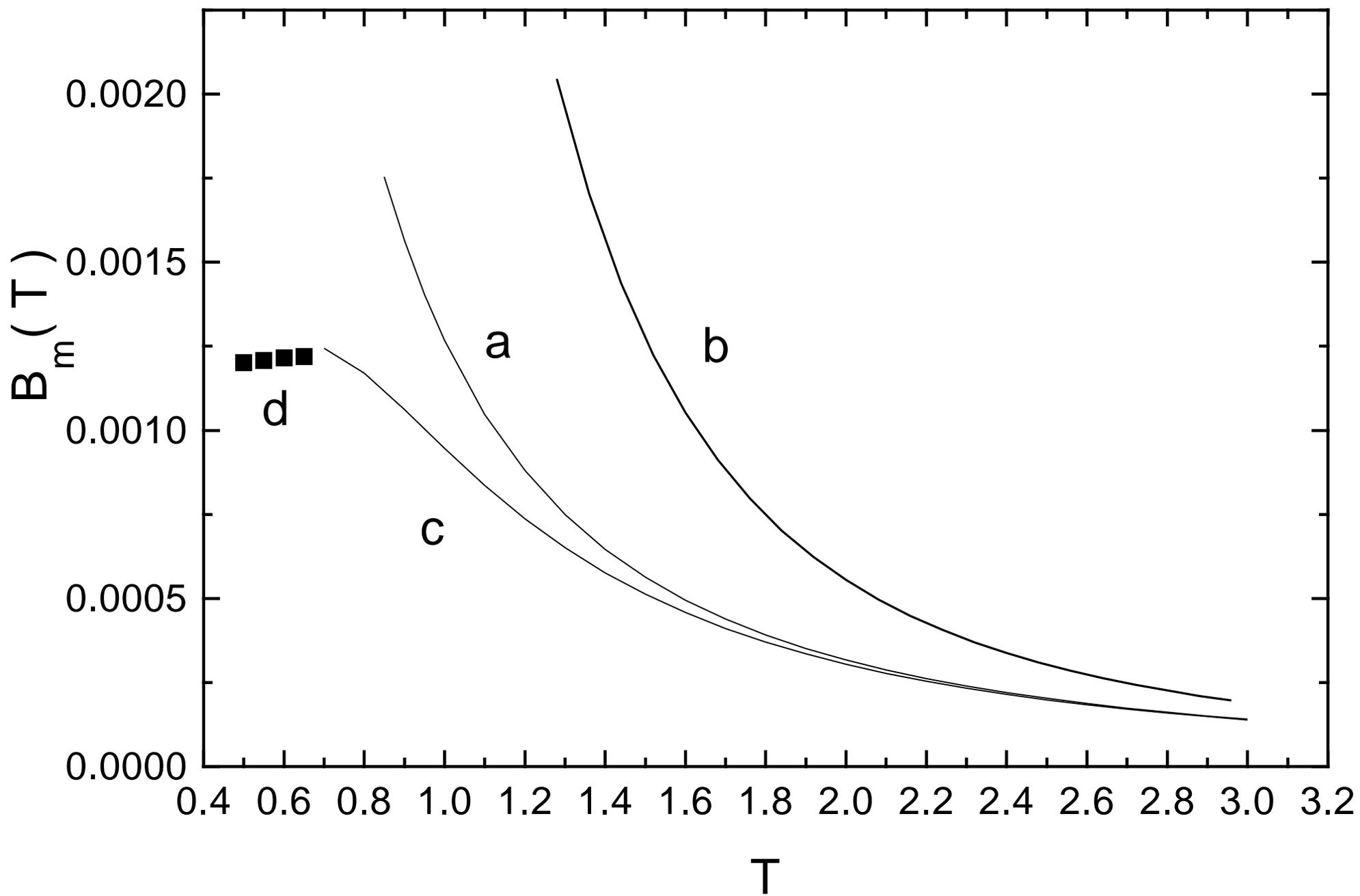
\end{document}